\RequirePackage{lineno}

\documentclass[aps,prl,twocolumn,groupedaddress,nofootinbib,showpacs]{revtex4}%

\usepackage{graphicx}
\usepackage{amsmath}
\usepackage{amssymb}
\usepackage{multirow}
\usepackage{bm}
\usepackage{color}

\usepackage[hypertex]{hyperref}

\def\be{\begin{linenomath*}\begin{equation}}
\def\ee{\end{equation}\end{linenomath*}}
\def\ba{\begin{linenomath*}\begin{eqnarray}}
\def\ea{\end{eqnarray}\end{linenomath*}}

\def\jpsi{{J/\psi}}
\def\moss{{\langle\mathcal{O}^\jpsi(\bigl.^3\hspace{-1mm}S_1^{[1]})\rangle}}
\def\mopa{{\langle\mathcal{O}^\jpsi(\bigl.^1\hspace{-1mm}S_0^{[8]})\rangle}}
\def\mopb{{\langle\mathcal{O}^\jpsi(\bigl.^3\hspace{-1mm}S_1^{[8]})\rangle}}
\def\mopc{{\langle\mathcal{O}^\jpsi(\bigl.^3\hspace{-1mm}P_0^{[8]})\rangle}}

\def\mossetac{{\langle\mathcal{O}^{\eta_c}(\bigl.^1\hspace{-1mm}S_0^{[1]})\rangle}}
\def\mopaetac{{\langle\mathcal{O}^{\eta_c}(\bigl.^3\hspace{-1mm}S_1^{[8]})\rangle}}
\def\mopbetac{{\langle\mathcal{O}^{\eta_c}(\bigl.^1\hspace{-1mm}S_0^{[8]})\rangle}}
\def\mopcetac{{\langle\mathcal{O}^{\eta_c}(\bigl.^1\hspace{-1mm}P_1^{[8]})\rangle}}

\def\mosochic{{\langle\mathcal{O}^{\chi_{c0}}(\bigl.^3\hspace{-1mm}S_1^{[8]})\rangle}}
\def\mopschic{{\langle\mathcal{O}^{\chi_{c0}}(\bigl.^3\hspace{-1mm}P_0^{[1]})\rangle}}
\def\mosohc{{\langle\mathcal{O}^{h_c}(\bigl.^1\hspace{-1mm}S_0^{[8]})\rangle}}
\def\mopshc{{\langle\mathcal{O}^{h_c}(\bigl.^1\hspace{-1mm}P_1^{[1]})\rangle}}

\def\sps{{\bigl.^1\hspace{-1mm}S^{[8]}_0}}
\def\so{{\bigl.^3\hspace{-1mm}S^{[8]}_1}}

\def\pj{{\bigl.^3\hspace{-1mm}P^{[8]}_J}}
\def\p0{{\bigl.^3\hspace{-1mm}P^{[8]}_0}}

\def\gev{\text{GeV}}

\begin{document}

% Use the \preprint command to place your local institutional report
% number in the upper righthand corner of the title page in preprint mode.
% Multiple \preprint commands are allowed.
% Use the 'preprintnumbers' class option to override journal defaults
% to display numbers if necessary
%\preprint{}

%Title of paper

\title{\Large $\eta_c$ production at LHC and implications for the understanding of $J/\psi$ production}
\author{
 Hao Han$^a$, Yan-Qing Ma$^{b,c}$, Ce Meng$^a$, Hua-Sheng Shao$^{a,d}$, Kuang-Ta Chao$^{a,c,e}$}
%\footnote{Email: ktchao@pku.edu.cn}}

\affiliation{ {\footnotesize (a)~School of Physics and State Key
Laboratory of Nuclear Physics and Technology, Peking University,
Beijing 100871, China}\\
{\footnotesize (b)~Maryland Center for Fundamental Physics, University of Maryland, College Park, Maryland 20742, USA}\\
 {\footnotesize (c)~Center for High Energy physics, Peking
University, Beijing 100871, China}\\
 {\footnotesize (d)~Physics Department, Theory Unit, CERN, CH-1211 Geneva 23, Switzerland}\\
 {\footnotesize (e)Collaborative Innovation Center of Quantum Matter, Beijing 100871, China}
}
%\date{\today}

%%%%%%%%%%%%%%%%%%%%%%%%%%%%%%%%%%%%%%%%%%%%%%%%%%%%%%%%%%%%%%%%%%%%%%%%%%%%%%
\begin{abstract}
We present a complete evaluation for the prompt $\eta_c$ production at the LHC at next-to-leading order in $\alpha_s$ in nonrelativistic QCD. By assuming heavy quark spin symmetry, the recently observed $\eta_c$ production data by LHCb results in a very strong constraint on the upper bound of the color-octet long distance matrix element $\mopa$ of $J/\psi$. We find this upper bound is consistent with our previous study of the $J/\psi$ yield and polarization and can give good descriptions for the measurements, but inconsistent with some other theoretical estimates. This may provide important information for understanding the nonrelativistic QCD factorization formulism.
%The recently observed $\eta_c$ production by LHCb may provide a crucial test for the charmonium production mechanism in nonrelativistic QCD (NRQCD). We present a complete evaluation for the prompt $\eta_c$ production at the LHC at next-to-leading order in $\alpha_s$ in NRQCD. By assuming heavy quark spin symmetry, with the LHCb data the study of $\eta_c$ production results in a very strong constraint on the upper bound of the color-octet long distance matrix element $\mopa$ of $J/\psi$. We find this upper bound is consistent with our previous study of the $J/\psi$ yield and polarization and can give good descriptions for the measurements, but inconsistent with many other theoretical estimates. This may provide important information for understanding the NRQCD factorization formulism.
\end{abstract}

%%%%%%%%%%%%%%%%%%%%%%%%%%%%%%%%%%%%%%%%%%%%%%%%%%%%%%%%%%%%%%%%%%%%%%%%%%%%%%
% insert suggested PACS numbers in braces on next line
% 13.66.Bc   Hadron production in e-e+ interactions
% 12.38.Bx   Perturbative calculations
% 13.60.Le   mesons production by photons and leptons
% 13.25.Gv   decays of hadronic
% 14.40.Pq    Heavy quarkonia
%%%%%%%%%%%%%%%%%%%%%%%%%%%%%%%%%%%%%%%%%%%%%%%%%%%%%%%%%%%%%%%%%%%%%%%%%%%%%%
% insert suggested keywords - APS authors don't need to do this
%\keywords{}

%%%%%%%%%%%%%%%%%%%%%%%%%%%%%%%%%%%%%%%%%%%%%%%%%%%%%%%%%%%%%%%%%%%%%%%%%%%%%%
%\maketitle must follow title, authors, abstract, \pacs, and \keywords
\maketitle

%%%%%%%%%%%%%%%%%%%%%%%%%%%%%%%%%%%%%%%%%%%%%%%%%%%%%%%%%%%%%%%%%%%%%%%%%%%%%%
% body of paper here - Use proper section commands
% References should be done using the \cite, \ref, and \label commands

{\it Introduction.}---
%The production of heavy quarkonium at hardron colliders is an ideal process to study QCD %at the interplay of perturbative and nonperturbative domains.
Significant improvements for understanding the heavy quarkonium production mechanism have been achieved in recent years. While abundant data of prompt heavy quarkonium production are accumulated at the LHC, one of the main theoretical improvements is to understand the yields and polarizations by including the higher order QCD effects in the framework of nonrelativistic QCD (NRQCD) factorization~\cite{Bodwin:1994jh}, where the inclusive production cross section of a quakonium state $\mathcal{Q}$ in $pp$ collisions can be expressed as
%%%%%%%%%%%%%%%%%%%%%%%%%%%%%%%%%
\ba\label{NRQCD}
d\sigma_{pp\to\mathcal{Q}+X}&=&\sum_{n}d{\hat{\sigma}}_{pp\to
Q\bar{Q}[n]+X}\langle\mathcal{O}^{\mathcal{Q}}(n)\rangle.
\ea
%%%%%%%%%%%%%%%%%%%%%%%%%%%%%%%%%
Here $d{\hat{\sigma}}_{pp\to Q\bar{Q}[n]+X}$ are the short-distance coefficients (SDCs) for producing a heavy quark pair $Q\bar{Q}$ with
quantum number $n$, and $\langle\mathcal{O}^{\mathcal{Q}}(n)\rangle$
are the long-distance matrix elements (LDMEs) for $\mathcal{Q}$. The SDCs can be computed in perturbative QCD as partonic cross sections convoluted with parton distributions. At large transverse momentum $p_T$,
%(the transverse momentum of $\mathcal{Q}$)
they behave roughly as some powers of $1/p_T$. The nonperturbative LDMEs can be arranged as a series in powers of $v$ (the relative velocity of the heavy quark $Q$ and antiquark $\bar{Q}$
%in the rest frame of $\mathcal{Q}$
)\cite{Bodwin:1994jh}. E.g., for the $J/\psi$ production the sum over $n$ can be truncated at order $v^4$, and the LDMEs are $\moss, \mopa,
\mopb$ and $\mopc$.
%%%%%%%%%%%%%%%%%%%%%%%%%%%%%%%%%
%\ba\label{LDMEs-Jpsi}
%\moss\sim\mathcal{O}(1),\mopa\sim\mathcal{O}(v^3),\nonumber\\
%\mopb\sim\mathcal{O}(v^4),\mopc\sim\mathcal{O}(v^4).\
%\ea
%%%%%%%%%%%%%%%%%%%%%%%%%%%%%%%%%SDCs relevant to the LDMEs in (\ref{LDMEs-Jpsi}) were calculated

In the past few years, complete next-to-leading order (NLO) QCD corrections for the $\jpsi$ hadroproduction have been calculated by three groups independently\cite{Ma:2010yw,Butenschoen:2010rq,Gong:2012ug}. Though the three groups obtained consistent SDCs, they had different philosophies on extracting the color-octet (CO) LDMEs, then got different CO LDMEs, and gave completely different predictions/descripctions for the polarization of prompt $\jpsi$. Specifically, our group found that the $\jpsi$ polarization can be explained by NLO NRQCD \cite{Chao:2012iv}, whereas the other two groups concluded that NRQCD can not explain the polarization data\cite{Butenschoen:2012px,Gong:2012ug}. More recently, two other groups performed independent fits for the CO LDMEs\cite{Bodwin:2014gia,Faccioli:2014cqa}, and concluded that the $\jpsi$ polarization can be understood by a $\sps$ channel dominance mechanism, which was first proposed as one possibility to explain the $\jpsi$ polarization in Ref.\cite{Ma:2010yw} and reemphasized in Ref.\cite{Chao:2012iv}.

To further test the mechanism of quarkonium production
%and to clarify the situation mentioned above,
it is crucial to have more measurements. Recently, for the first time the LHCb Collaboration  measured the $p_T$ differential cross section of  prompt $\eta_c$ production via $\eta_c\to p\bar p$ \cite{Aaij:2014bga}. This measurement is not only significant for $\eta_c$ production, but also provides important information for $\jpsi$ production via heavy quark spin symmetry (HQSS)~\cite{Bodwin:1994jh}. Leading order study of $\eta_c$ hadroproduction can be found in Refs.~\cite{Mathews:1998nk,Likhoded:2014fta} and references therein. In this letter, we will study the $\eta_c$ hadroproduction at NLO in $\alpha_s$ within the framework of NRQCD factorization. With HQSS, we find that the $\eta_c$ production data are compatible with our previous studies\cite{Ma:2010yw,Chao:2012iv} and may provide a further constraint on the possible values of $\jpsi$ LDMEs.

{\it Relationship between $\jpsi$ and $\eta_c$ production.}--- Let's first explain why $\eta_c$ production can provide important clues to $\jpsi$ production.

For the $\jpsi$ production,  with a relative large $p_T$ cutoff ($p_T>7~$GeV), our group found that~\cite{Ma:2010yw} only two linear combinations of the three CO LDMEs can be well constrained by fitting the CDF data~\cite{Acosta:2004yw} of the yields of $J/\psi$ production, which gives
%%%%%%%%%%%%%%%%%%%%%%%%%%%%%%%%%%%%%
\begin{eqnarray}\label{M0M1}
M_0 &=& \langle\mathcal{O}^{J/\psi}(^1\!S_0^{[8]})\rangle + r_0 \langle\mathcal{O}^{J/\psi}(^3\!P_0^{[8]})\rangle/m_c^2,\nonumber\\
M_1 &=& \langle\mathcal{O}^{J/\psi}(^3\!S_1^{[8]})\rangle + r_1 \langle\mathcal{O}^{J/\psi}(^3\!P_0^{[8]})\rangle/m_c^2,
\end{eqnarray}
%%%%%%%%%%%%%%%%%%%%%%%%%%%%%%%%%%%%%
where $r_0 = 3.9$ and $r_1 = -0.56$ for the CDF window, and the corresponding values are $M_0=(7.4\pm1.9)\times 10^{-2}~\mbox{GeV}^3$ and $M_1=(0.05\pm0.02)\times 10^{-2}~\mbox{GeV}^3$. Roughly speaking, the SDCs for the LDMEs $M_0$ and $M_1$ defined in (\ref{M0M1}) have mainly $p_T^{-6}$ and $p_T^{-4}$ behaviors~\cite{Ma:2010yw}, respectively. These two $p_T$ behaviors dominate the $\jpsi$ production in the region $p_T>7~$GeV. The coefficients $r_0$ and $r_1$ change slightly with the rapidity interval but almost not change with the center-of-mass energy $\sqrt{S}$ (see Table I in Ref.~\cite{Ma:2010jj}). Thus, the CMS yield data~\cite{Chatrchyan:2011kc} for $J/\psi$ production can be also well described by the same LDMEs in Eq.~(\ref{M0M1})~\cite{Ma:2010yw,Shao:2014yta}. Importantly, we further found that~\cite{Chao:2012iv} the transversely polarized cross section for direct $J/\psi$ production at NLO is almost proportional to the combined LDME
\begin{eqnarray}\label{M1prime}
M_1^{\prime}=\langle\mathcal{O}^{J/\psi}(^3\!S_1^{[8]})\rangle - 0.52\langle\mathcal{O}^{J/\psi}(^3\!P_0^{[8]})\rangle/m_c^2
\end{eqnarray}
for the CDF and CMS window, which is very close to the $M_1$ in (\ref{M0M1}). Since the value of $M_1$ is extremely small, much smaller than that of $M_0$ in Eq.~(\ref{M0M1}), one can expect that the polarizations will be dominated by $M_0$ at least in the intermediate $p_T$ region, which tends to give unpolarized results~\cite{Chao:2012iv}. We emphasize here that the above expectation is independent of the exact values of the three CO LDMEs of $J/\psi$, as long as $M_0$ and $M_1$ are fixed by Eq.~(\ref{M0M1}). This can be seen from the fact that, by varying the value of $\mopa$ in Table I of Ref.~\cite{Chao:2012iv},
the resulted polarizations are similar and basically unpolarized~\cite{Chao:2012iv}. Based on Eq.~(\ref{M0M1}), assuming all CO LDMEs to be positive, we updated our results for the polarization of direct $J/\psi$ production together with that of the feeddown contributions from $\chi_c$ and $\psi(2S)$ in Ref.~\cite{Shao:2014yta}, which are roughly consistent with the LHC data.
%Needless to say, further constraints on the CO LDMEs from data are the main way to test %the above solution for the $\jpsi$ polarization.

The cross section for $\eta_c$ production can also be expressed as Eq.~(\ref{NRQCD}). Similar to the case for $J/\psi$, four LDMEs are needed up to relative order $v^4$ for the direct $\eta_c$ production, which are $\mossetac,\mopaetac,\mopbetac$ and $\mopcetac$. The dominant feeddown contribution through $h_c\to\eta_c\gamma$ introduces two other LDMEs at relative order $v^2$: $\mopshc$ and $\mosohc$. Superficially, it appears that six channels would be involved in the fit to data, but in fact, some of them are not important. The relative importance of these channels should depend on the power counting both in $v$ and in $\delta=m_{\mathcal{Q}}/p_T$, where $m_{\mathcal{Q}}$ is the mass of the charmonium. The powers of $v$ can be estimated by the velocity scaling rules~\cite{Bodwin:1994jh}. The powers of $\delta$ can be determined by QCD factorization for quarkonium production \cite{Kang:2014tta}, which shows that all channels have a leading power (LP), $p_T^{-4}$, component at the current order in $\alpha_s$. However, because of the relative importance of next-to-leading power (NLP), $p_T^{-6}$, contributions for some channels \cite{Ma:2014svb},  they will behave almost as $p_T^{-6}$ within a large range of $p_T$.

%%%%%%%%%%%%%%%%%%%%%%%%%%%%%%%%%%%%%
\begin{table}[!htp]
\begin{tabular}{{ccccccc}}
\hline\hline \itshape
~$n=$~&~$^1\hspace{-1mm}S_0^{[1]}$~&~$^3\hspace{-1mm}S_1^{[8]}$~&~$^1\hspace{-1mm}P_1^{[8]}$~&~$^1\hspace{-1mm}S_0^{[8]}$~&~$^1\hspace{-1mm}S_0^{[8]}(h_c)$~&~$^1\hspace{-1mm}P_1^{[1]}(h_c)$~
\\\hline  & $v^0\delta^6$ &  $v^{3}\delta^4$ &  $v^4\delta^6$ &  $v^4\delta^6$  &  $v^2\delta^6$  &  $v^2\delta^6$
\\\hline\hline
\end{tabular}
\caption{\label{PCRs} The power counting results in double expansions in powers of $v$ and $\delta=m_{\mathcal{Q}}/p_T$ for different channels $n$ relevant to the prompt $\eta_c$ production for the LHCb window~\cite{Aaij:2014bga}.}
\end{table}
%%%%%%%%%%%%%%%%%%%%%%%%%%%%%%%%%%%%%

Especially, for the LHCb window, e.g., $6.5~\mbox{GeV}<p_T<14~\mbox{GeV}$~\cite{Aaij:2014bga}, effectively only the $\so$ channel behaves as $p_T^{-4}$, while all other channels behave as $p_T^{-6}$, as shown in Table~\ref{PCRs}. As a result, only the $^1\hspace{-1mm}S_0^{[1]}$ and the $^3\hspace{-1mm}S_1^{[8]}$ channels give the leading contributions in the combined power counting. By further applying the HQSS relation~\cite{Bodwin:1994jh}
\begin{eqnarray}\label{HQSS-2}
\mopaetac &\approx& \mopa,
\end{eqnarray}
which is valid up to $O(v^2)$ corrections, one may expect that the LDME $\mopa$ will be determined by the study of $\eta_c$ production. This will give the third independent constraint on the three CO LDMEs for $J/\psi$ production other than those given in Eq.~(\ref{M0M1}).

{\it The $\eta_c$ production}---Let us proceed to study the $\eta_c$ production numerically. We use the CTEQ6M PDFs~\cite{Whalley:2005nh} for NLO calculations, and use HELAC-Onia\cite{Shao:2012iz} to calculate the hard non-collinear part of real correction. The charm-quark mass is set to be $m_c = 1.5$~GeV, the renormalization, factorization, and NRQCD scales are $\mu_r = \mu_f = \sqrt{p_T^2+4m_c^2}$ and $\mu_{\Lambda} = m_c$. Thanks to HQSS, the color-singlet (CS) LDMEs for both $\jpsi$ and $\eta_c$ can be estimated by the potential model \cite{Eichten:1995ch},
%%%%%%%%%%%%%%%%%%%%%%%%%%%%%%%%%%%%%
\begin{eqnarray}\label{HQSS-1}
\mossetac &=& \moss/3=0.39~\mbox{GeV}^3.
\end{eqnarray}
%%%%%%%%%%%%%%%%%%%%%%%%%%%%%%%%%%%%%
The theoretical uncertainties by varying $m_c$, $\mu_f$ and $\mu_r$ have been studied thoroughly in our earlier publications \cite{Ma:2010vd,Ma:2010yw,Ma:2010jj,Chao:2012iv}, where one found that the uncertainties can be estimated by a systematical error of about 30\%.

\begin{figure}
\includegraphics[scale=0.35]{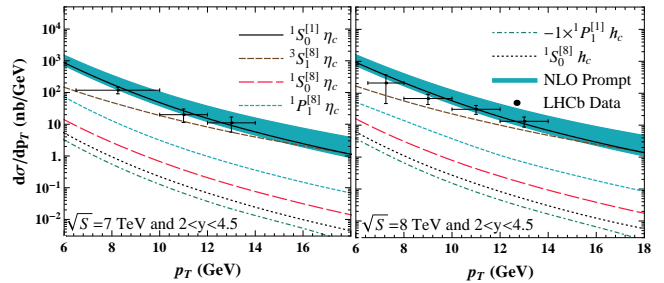}
\caption{The differential cross sections of prompt $\eta_c$ production at $\sqrt{S}=7$~TeV (left) and 8~TeV (right) for the LHCb window. The data are taken from Ref.~\cite{Aaij:2014bga}. See text for details.
%The CS contributions are denoted by the solid lines. The methods to estimate the contributions from other channels have been explained in the text. The theoretical errors indicated by the blue band are mainly from the uncertainties of the LDME $\mopaetac$ in Eq.~(\ref{UpperLimit}).
}
 \label{eatc-total}
\end{figure}

As mentioned above, only the channels $^1\hspace{-1mm}S_0^{[1]}$ and $^3\hspace{-1mm}S_1^{[8]}$ are essential to account for the $\eta_c$ production in the LHCb window. However, with the fixed value of $\mossetac$ in Eq.~(\ref{HQSS-1}), we find that the LHCb data are almost saturated by the contribution from the CS channel, which is denoted by the solid lines in Fig.~\ref{eatc-total}. Similar results have been found in Ref.~\cite{Likhoded:2014fta} with only the LO SDCs and a relative smaller CS LDME. The similarity is mainly caused by that the NLO calculation gives only a modest correction factor for $^1\hspace{-1mm}S_0^{[1]}$ channel. Therefore, the saturation can hardly be avoided if one choose the CS LDME as large as that in Eq.~(\ref{HQSS-1}).

However, the above result does not mean that there is no contribution from the $^3\hspace{-1mm}S_1^{[8]}$ channel. On the one hand, although there are large uncertainties of the data, one can roughly find in Fig.~\ref{eatc-total} that the slope of data is different from the contribution of $^1\hspace{-1mm}S_0^{[1]}$ channel itself. On the other hand, the value in  Eq.~(\ref{HQSS-1}) is not exact, but with at least an uncertainty of order $v^2\sim0.3$ because of modeling of potential, relativistic corrections, HQSS broken, and so on. These uncertainties may leave some room for $^3\hspace{-1mm}S_1^{[8]}$ channel to contribute.

Unfortunately, it is very hard at present to determine the exact value of $\mopaetac$ due to the large uncertainties from both data and theory. But we may give a safe upper bound for $\mopaetac$ by letting the data be saturated by the $^3\hspace{-1mm}S_1^{[8]}$ channel only, which gives $\mopaetac=(1.46\pm0.20)\times10^{-2}~\mbox{GeV}^3$. Since the value of $\mopaetac$ is sufficient amplified, we choose the central value as the upper bound for the LDME. To give a lower bound, we assume the $\mopaetac$ to be positive \cite{Shao:2014yta}, which should be acceptable due to the following reason. Since the renormalization dependence of  LDME $\mopaetac$ is at higher order in $v^2$, $\mopaetac$ can be approximated as the probability for the $c\bar{c}$ pair in $^3\hspace{-1mm}S_1^{[8]}$ configuration to evolve into $\eta_c$, which should be positive in general. We then get
%the range of $\mopaetac$,
%%%%%%%%%%%%%%%%%%%%%%%%%%%%%%%%%
\ba\label{UpperLimit}
0<\mopaetac<1.46\times10^{-2}~\mbox{GeV}^3.
\ea
%%%%%%%%%%%%%%%%%%%%%%%%%%%%%%%%%
By using the HQSS relation , the result in Eq.~(\ref{UpperLimit}) can be viewed as another constraint on the three CO LDMEs for $J/\psi$ other than Eq.~(\ref{M0M1}). We thus constrain  all three CO LDMEs of $J/\psi$ into a finite range.

As a feedback, the other two CO LDMEs for direct $\eta_c$ production can be estimated by the HQSS relations~\cite{Bodwin:1994jh}: $\mopbetac = \mopb/3$ and $\mopcetac = 3\mopc$.
%%%%%%%%%%%%%%%%%%%%%%%%%%%%%%%%%%%%%%
%\begin{eqnarray}\label{HQSS-3}
%\mopbetac &=& \mopb/3,\nonumber\\
%\mopcetac &=& 3\mopc.
%\end{eqnarray}
%%%%%%%%%%%%%%%%%%%%%%%%%%%%%%%%%%%%%%
As for the feeddown contribution through $h_c\to\eta_c\gamma$, the two relevant LDMEs can be estimated again by the HQSS relations: $\mosohc = 3\mosochic$ and $\mopshc = 3\mopschic$,
%%%%%%%%%%%%%%%%%%%%%%%%%%%%%%%%%%%%%%
%\begin{eqnarray}\label{HQSS-4}
%\mosohc &=& 3\mosochic,\nonumber\\
%\mopshc &=& 3\mopschic,
%\end{eqnarray}
%%%%%%%%%%%%%%%%%%%%%%%%%%%%%%%%%%%%%%
where the LDMEs for $\chi_{c0}$ have been determined in Ref.~\cite{Ma:2010vd,Shao:2014fca}. Combining the LDMEs estimated by the relations and the the SDCs calculated up to  NLO in $\alpha_s$, we show the sizes of the contributions from these channels in Fig.~\ref{eatc-total}, all of which are smaller than that for the CS channel by about one or two orders of magnitude as expected. Thus, the upper bound of the value of $\mopaetac$ given in Eq.~(\ref{UpperLimit}) will not be changed even these new contributions are taken into account. In addition, to provide an order of magnitude estimation of the contributions from the $^3\hspace{-1mm}S_1^{[8]}$ channel, we use a half of the upper bound of $\mopaetac$ as its input, and the results are shown as the middle-width dashed lines in Fig.~\ref{eatc-total}. The theoretical errors, which are indicated by the blue band in Fig.~\ref{eatc-total}, are mainly from the uncertainties of the LDME $\mopaetac$ in Eq.~(\ref{UpperLimit}).

%%%%%%%%%%%%%%%%%%%%%%%%%%%%%%%%%%%%%
%%%%%%%%%%%%%%%%%%%%%%%%%%%%%%%%%%%%%

%%%%%%%%%%%%%%%%%%%%%%%%%%%%%%%%%%%%%%
%\begin{figure}
%\includegraphics[width=4.5cm]{LHCb7TeV.eps}\hspace{-1cm}
%\includegraphics[width=4.5cm]{LHCb8TeV.eps}
%\caption{The differential cross sections of prompt $\eta_c$ production at $\sqrt{S}=7$~TeV (left) and 8~TeV (right) for the LHCb window. The data are taken from Ref.~\cite{Aaij:2014bga}. See text for details.
%%The CS contributions are denoted by the solid lines. The methods to estimate the contributions from other channels have been explained in the text. The theoretical errors indicated by the blue band are mainly from the uncertainties of the LDME $\mopaetac$ in Eq.~(\ref{UpperLimit}).
%}.
% \label{eatc-total}
%\end{figure}
%%%%%%%%%%%%%%%%%%%%%%%%%%%%%%%%%%%%%%

\begin{figure}[!tbhp]
\begin{center}
\includegraphics*[scale=0.335]{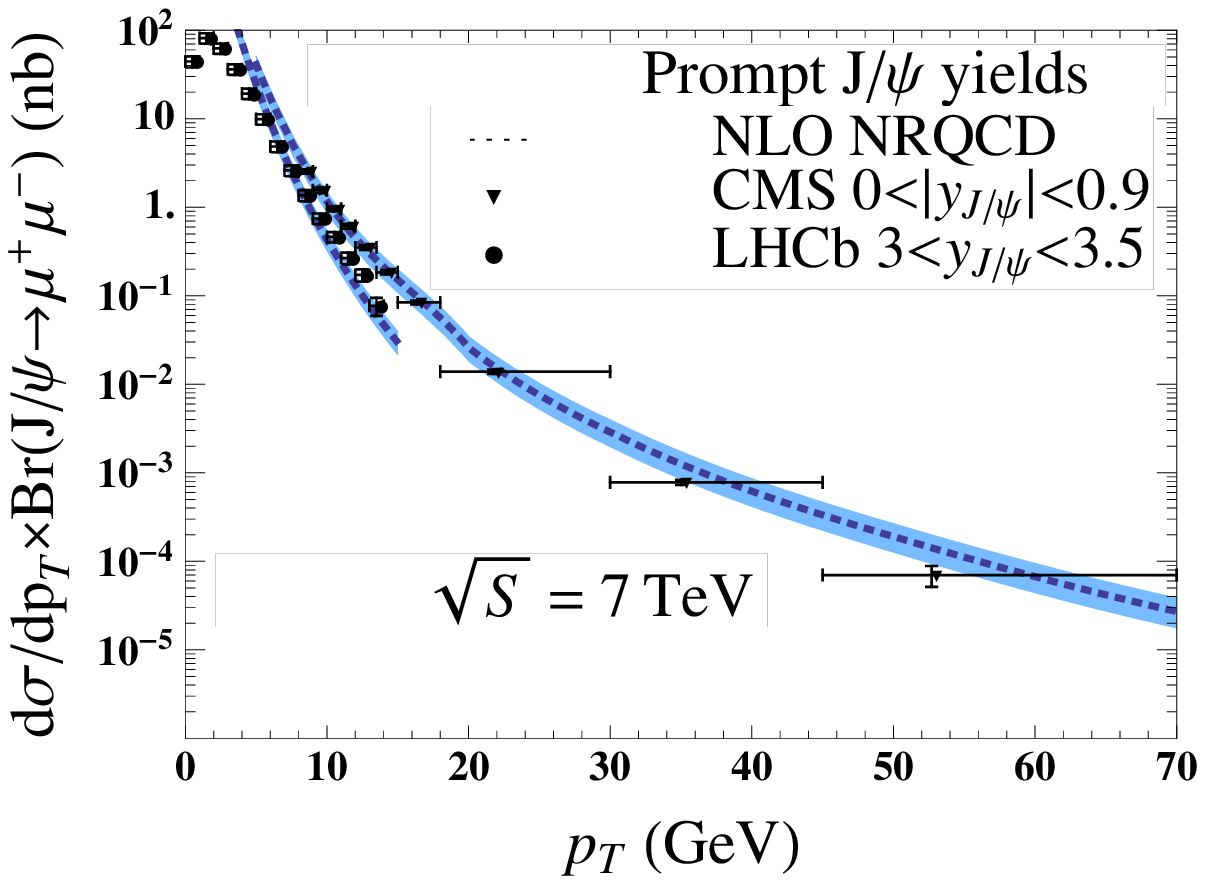}
\includegraphics*[scale=0.335]{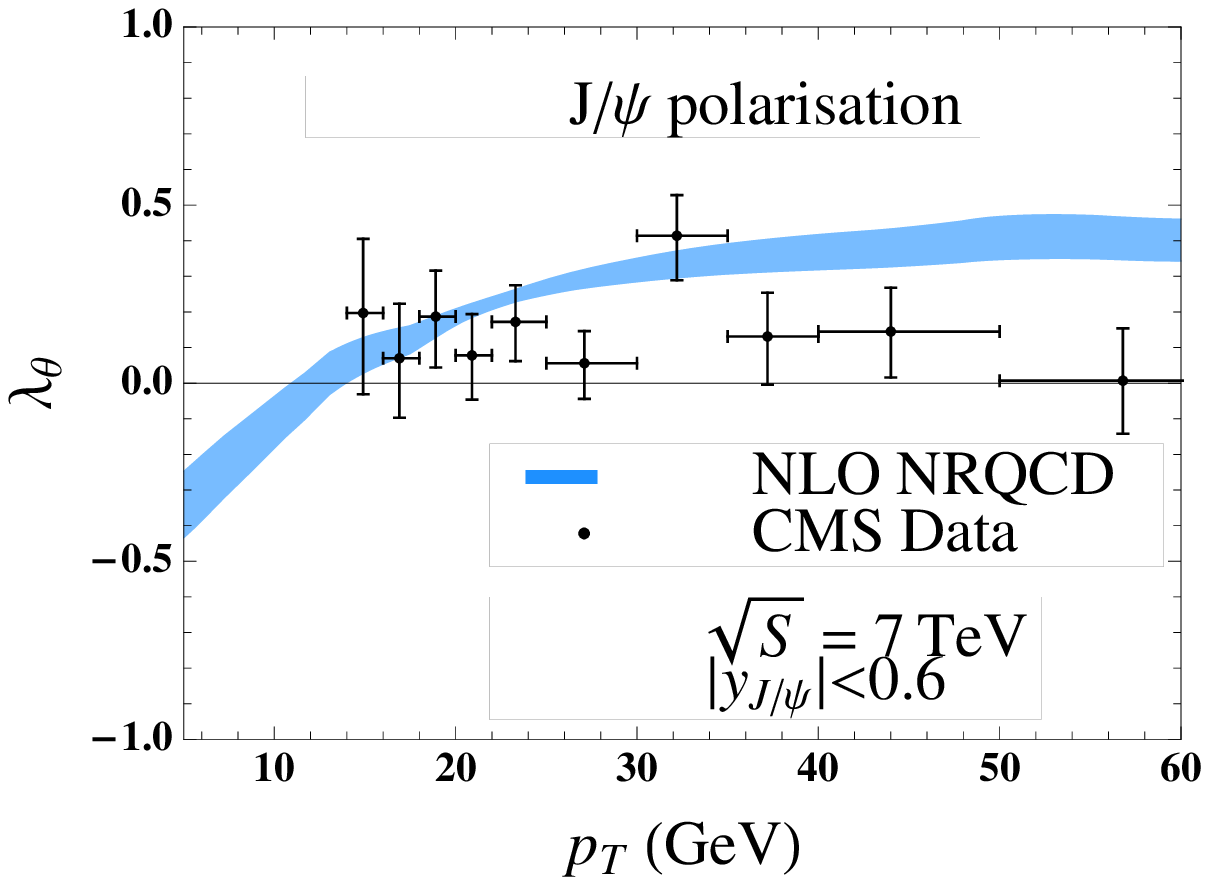}
\includegraphics*[scale=0.335]{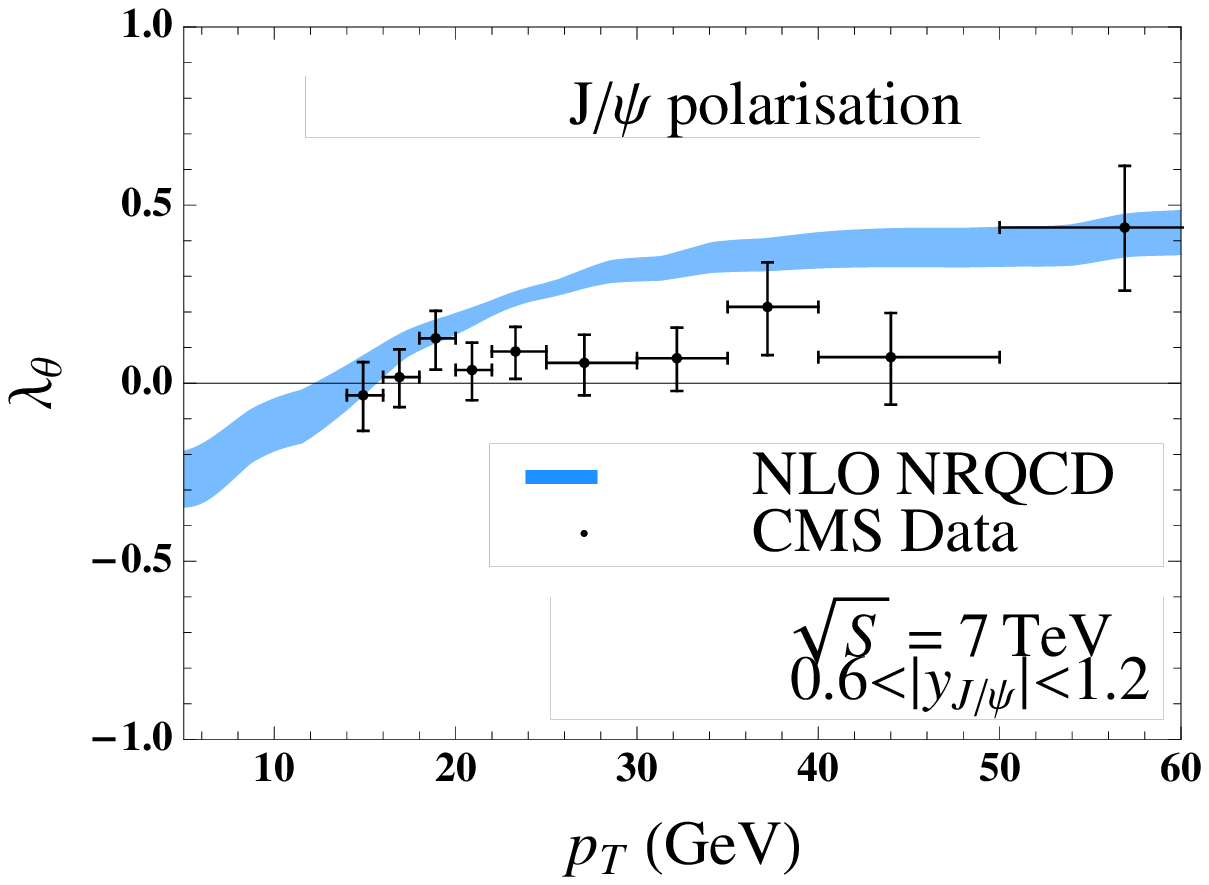}
\includegraphics*[scale=0.335]{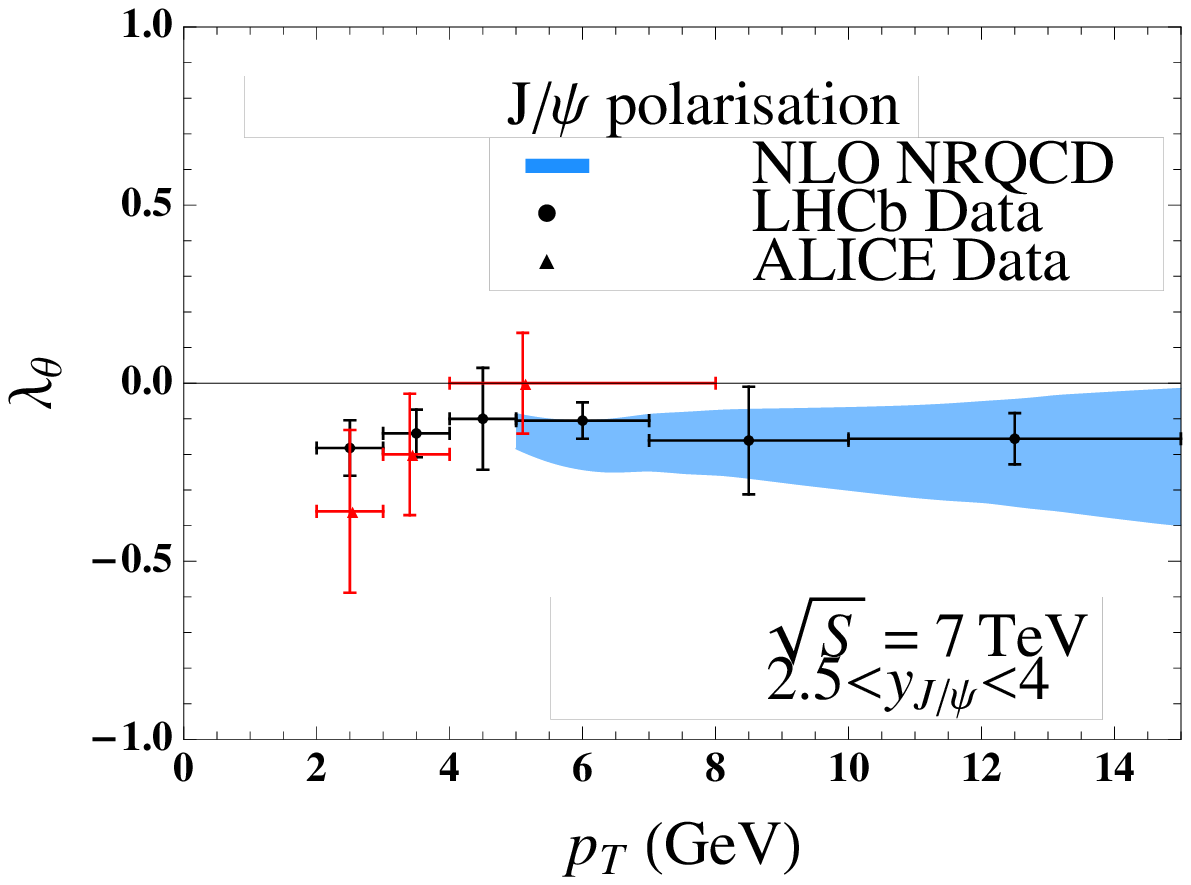}
\caption{Predictions for prompt $\jpsi$ production. Theoretical parameters are constrained by $\jpsi$ yield data at CDF \cite{Acosta:2004yw} as well as $\eta_c$ yield data at LHCb \cite{Aaij:2014bga} along with HQSS. Data are taken from CMS \cite{Chatrchyan:2011kc,Chatrchyan:2013cla}, LHCb \cite{Aaij:2011jh,Aaij:2013nlm} and ALICE \cite{Abelev:2011md}.}
 \label{fig:polarjpsi}
 \end{center}
\end{figure}

{\it Indications on the $\jpsi$ production.}---Let us go back to the problem of the $J/\psi$ production. Since the three CO LDMEs for $\jpsi$ can be constrained better by Eqs. (\ref{M0M1}) and (\ref{UpperLimit}) using the HQSS relation Eq.~(\ref{HQSS-2}), we update our predictions for both yields and polarizations of $J/\psi$ prompt production, which are shown in Fig.~\ref{fig:polarjpsi}. The details for these calculations have been explained in Ref.~\cite{Shao:2014yta}. Compared with the old results given in Ref.~\cite{Shao:2014yta}, the new predictions for the CMS window are almost unchanged. This is because, for the CMS window, the prediction for yield is only sensitive to the LDMEs $M_0$ and $M_1$ defined in Eq.~(\ref{M0M1}), and that for polarizations is only sensitive to $M_1^{\prime}$, which is given in (\ref{M1prime}) and very close to $M_1$ as mentioned above. Thus, these predictions can hardly be influenced by the extra constraint in Eqs.~(\ref{HQSS-2}) and (\ref{UpperLimit}). On the other hand, since $r_1$ in the forward rapidity interval is smaller than that in the central rapidity interval~\cite{Ma:2010jj}, the relative large and positive $\mopc$, which is indicated by Eqs.~(\ref{M0M1}), (\ref{HQSS-2}) and (\ref{UpperLimit}), will imply that the transversely polarized component of the cross section should be further reduced in the forward rapidity interval comparing with that in the central one. As a result, our new prediction of the polarization for the LHCb window with the new constraint Eqs.~(\ref{HQSS-2}) and (\ref{UpperLimit}) tends to be more longitudinally polarized. This slightly improves the consistency between the theoretical predictions and the experimental measurements  compared with that in Ref.~\cite{Shao:2014yta}. We list the $\chi^2/d.o.f.$ values for the polarization data: 13/10 and 22/10 for the CMS data with $0<|y|<0.6$ and $0.6<|y|<1.2$ respectively and 1.2/2 for the LHCb data. Although the agreement between our predictions and the CMS polarization data in Fig.~\ref{fig:polarjpsi} is not very good, it is tolerable considering the large experimental and theoretical uncertainties in this stage. In particular, we note that the current CMS data
still suffer from large statistical fluctuations, such as in the
last bins in |y| < 0.6 and 0.6 < |y| < 1.2.

%%%%%%%%%%%%%%%%%%%%%%%%%%%%%%%%%%%%
%%%%%%%%%%%%%%%%%%%%%%%%%%%%%%%%%%%%

%%%%%%%%%%%%%%%%%%%%%%%%%%%%%%%%%%%%%
%\begin{figure}[!tbhp]
%\begin{center}
%\includegraphics*[scale=0.1]{plots.eps}
%\caption{Predictions for prompt $\jpsi$ production. Theoretical parameters are constrained by $\jpsi$ yield data at CDF \cite{Acosta:2004yw} as well as $\eta_c$ yield data at LHCb \cite{Aaij:2014bga} along with HQSS. Data are taken from CMS \cite{Chatrchyan:2011kc,Chatrchyan:2013cla}, LHCb \cite{Aaij:2011jh,Aaij:2013nlm} and ALICE \cite{Abelev:2011md}.}
% \label{fig:polarjpsi}
% \end{center}
%\end{figure}
%%%%%%%%%%%%%%%%%%%%%%%%%%%%%%%%%%%%%

The above calculations and analysis indicate that the new constraint Eq.~(\ref{UpperLimit}) can hardly change our previous conclusions on the $\jpsi$ production~\cite{Ma:2010yw,Chao:2012iv,Shao:2014yta}. One should note that the HQSS in Eq.~(\ref{HQSS-2}) could be violated up to relative order $v^2$.
But the violation at this level will not change our conclusion qualitatively.

However, the upper bound of $\mopa$ obtained in Eq.~\eqref{UpperLimit} along with HQSS disagree with many other NLO NRQCD fits in the literature \cite{Butenschoen:2010rq,Gong:2012ug,Bodwin:2014gia,Faccioli:2014cqa}. In Refs.~\cite{Butenschoen:2010rq,Gong:2012ug,Bodwin:2014gia}, the $\mopa$ is found to be well constrained, with the value $0.0304\pm0.0035~\gev^3$, $0.097\pm0.009~\gev^3$ and $0.099\pm0.022~\gev^3$, respectively. While in Ref.~\cite{Faccioli:2014cqa}, the authors argued that $\sps$ will dominate the $\jpsi$ production, and thus their $\mopa$ should be at least larger than $0.07~\gev^3$. As we discussed above, Eq.~\eqref{UpperLimit} gives a very safe upper bound for $\mopa$, so the contradiction with these NLO NRQCD fits may indicate that either HQSS is essentially broken or there are still some theoretical problems to be clarified, if the LHCb data\cite{Aaij:2014bga} are reliable.

Though both Refs.~\cite{Butenschoen:2010rq,Gong:2012ug,Faccioli:2014cqa} and our works~\cite{Ma:2010yw,Ma:2010jj,Shao:2014yta} are based on complete NLO NRQCD calculations, there are many differences in the fit procedures. E.g., the lower $p_T$ cutoff for experimental data is chosen to be $1\gev$ in Ref.~\cite{Butenschoen:2010rq} (for the photoproduction data), $7\gev$ in Refs.~\cite{Ma:2010yw,Gong:2012ug}, and $3$ times mass of $\jpsi$ in Ref.~\cite{Faccioli:2014cqa}; feeddown contributions are considered in Refs.~\cite{Ma:2010yw,Gong:2012ug}, but not considered in Ref.~\cite{Butenschoen:2010rq}. So further studies are needed to uncover the deep reason for the discrepancies of these fits and to explore the best value of lower $p_T$ cutoff for experimental data. It is interesting to compare the work of Ref.~\cite{Bodwin:2014gia} with ours. In addition to our complete NLO NRQCD results, the crucial element that Ref.~\cite{Bodwin:2014gia} includes is a partial LP contribution at next-to-next-to-leading order (NNLO) in $\alpha_s$. We conjecture that it is mainly this extra LP contribution that changes the theoretical curve of $\pj$ channel, and  results in a $\sps$ dominance conclusion in Ref.~\cite{Bodwin:2014gia}. If HQSS is good, a natural way to solve the contradiction could be that the NNLO correction for NLP contribution is also significant. It is needed to perform complete calculations for both the LP and NLP contributions to the same order in $\alpha_s$. Based on QCD factorization up to NLP \cite{Kang:2014tta} and the method to calculate the partonic hard part at NLP~\cite{Kang:2014pya}, the NNLO correction for NLP contribution may be achieved  soon. Then the validation of HQSS for charmonium production will be tested on a more rigorous base.

{\it Summary.}---%The recently observed $\eta_c$ production by LHCb\cite{Aaij:2014bga} may provide a crucial test for the quarkonium production mechanism in the NRQCD factorization formulism.  We compute the cross section of prompt $\eta_c$  production at the LHC at NLO in $\alpha_s$ in NRQCD.
Within NLO NRQCD, we demonstrate that only $^1\hspace{-1mm}S_0^{[1]}$ and $^3\hspace{-1mm}S_1^{[8]}$ channels are essential for the $\eta_c$  production at LHC. By comparing with the LHCb data~\cite{Aaij:2014bga}, we find the $\eta_c$  production tends to be saturated by contributions from the CS channel. This strongly constrains the CO LDME $\mopaetac$ for $\eta_c$, which can be related to $\mopa$ for $J/\psi$ by the HQSS relation in Eq.~(\ref{HQSS-2}). With the help of this new information, all three CO LDMEs of $\jpsi$ can be well constrained into a finite range.
%Then, we update our predictions of yield and polarization in $J/\psi$ prompt production, which are roughly the same as, but slightly better than the old ones~\cite{Ma:2010yw,Chao:2012iv,Shao:2014yta}.
We conclude that the prompt production of $\eta_c$ and $J/\psi$ can be understood in the same theoretical framework. Moreover, we find some previous works\cite{Butenschoen:2010rq,Gong:2012ug,Bodwin:2014gia,Faccioli:2014cqa}
%, where $\mopa$ is well constrained by fitting the data of $J/\psi$ production,
may overestimate the value of $\mopa$ unless HQSS is broken. All these studies on $\eta_c$ and $J/\psi$ may provide important information for understanding the mechanism of charmonium production.
%Further experimental and theoretical studies are needed to clarify the present issues on $\eta_c$ and $J/\psi$ hadroproduction.

{\it Note added.}---When our calculation was finished and the manuscript was being prepared for publication, an independent study of $\eta_c$ production in NLO NRQCD was reported in Ref.~\cite{Butenschoen:2014dra}. These authors conclude that with HQSS the $\eta_c$ data conflict with all NLO NRQCD fits to $\jpsi$ production data. This conclusion differs from ours, and is due to their using the values of the first row of Table I in Ref.~\cite{Chao:2012iv} but not that of the second and third rows of the same Table. Indeed, we emphasized in Ref.\cite{Chao:2012iv} that ``As the yield and polarization share a common parameter
space, and the yield can only constrain two linear
combinations of CO LDMEs, the combined fit of both
yield and polarization may also not constrain three independent
CO LDMEs stringently. In fact we find for a wide
range of given $\mopa$, one can fit both yield and polarization
reasonably well" and we showed in the second row of Table I that a possible value for $\mopa$ can be even as small as zero. We repeated the conclusion again in our most recent paper~\cite{Shao:2014yta} that only the two linear combinations of CO LDMEs in Eq.~\eqref{M0M1} can be well constrained.

%{\it Note added in production.}---When our paper was submitted for review, another independent study of $\eta_c$ production in NLO NRQCD was reported in Ref.~\cite{Zhang:2014ybe}. Their results are consistent with ours.

{\it Acknowledgments.}---We thank V. Belyaev, Z. Yang and S. Barsuk for helpful discussions on the experiments of $\eta_c$ production at the LHCb. This work was supported in part by the National Natural Science Foundation of China (No 11075002, No 11475005) and the National Key Basic Research Program of China (No 2015CB856700). Y.Q.M. is supported by the U.S. Department of Energy Office of Science, Office of Nuclear Physics under Award Number DE-FG02-93ER-40762.

%~~~~~~~~~~~~~~~~~~
%\begin{acknowledgments}

%\end{acknowledgments}

%%%%%%%%%%%%%%%%%%%%%%%%%%%%%%%%%%%%%%%%%%%%%%%%%%%%%%%%%%%%%%%%%%%%%%%%%%%%%%
% Create the reference section using BibTeX:
%\bibliography{paper}% Produces the bibliography via BibTeX.
%\bibliographystyle{utphysPhysRev}
%\bibliography{bibTex1.4}

\providecommand{\href}[2]{#2}\begingroup\raggedright\endgroup

\end{document}